\documentstyle[aps,prc,twocolumn,floats,psfig]{revtex}

\begin{document}

\draft

\title{Moments of inertia for multi-quasiparticle configurations}

\author{S.~Frauendorf$^{(1)}$, K.~Neerg{\aa}rd$^{(2)}$, J.~A. Sheikh$^{(3,4)}$
        and P.~M.~Walker$^{(3)}$}

\address{
$^{(1)}$Department of Physics, University of Notre Dame, Notre Dame,
        Indiana 46556, USA\\
$^{(2)}$N{\ae}stved Gymnasium og HF,
        Nyg{\aa}rdsvej 43, 4700 N{\ae}stved, Denmark\\
$^{(3)}$Department of Physics, University of Surrey, Guildford,
        Surrey GU2 5XH, UK\\
$^{(4)}$Tata Institute of Fundamental Research,
        Bombay, 400 005, India
}


\maketitle

\begin{abstract}

Tilted-axis cranking calculations have been performed for multi-quasiparticle
states in well deformed A$\approx$180 nuclei. In the limit of zero pairing, not
only are the calculated moments of inertia substantially smaller than for rigid
rotation, but also they are close to the experimental values. The moments of
inertia are found to be insensitive to dynamic pair correlations.

\end{abstract}

\pacs{PACS numbers: 21.10.Re, 21.30.Fe, 21.60.Ev, 27.70.+q}

\section{Introduction}\label{intr}

The moments of inertia of deformed atomic nuclei at low spins are a factor of
two or three smaller than the rigid-body value. The reduction is attributed to
the strong pair correlations, because nuclei in the ground state are in a
superfluid condensed state \cite{bm2}. Angular momentum is generated by either
rotating the deformed superfluid or by breaking Cooper pairs from this
condensed state. In order to reach high spins, an increasing number of Cooper
pairs are broken, which reduces and finally quenches the pair condensate. It
has been often stated that after this transition, the moments of inertia should
reach the rigid-body value~\cite{bm2,ring}. However, this conjecture is based
on the consideration of two special cases~\cite{bm2,ring}: (i) the limit of
large particle number, where the nuclear shell structure becomes unimportant;
and (ii) nucleons in a harmonic oscillator potential at its equilibrium
deformation.

Particularly for independent particles in a harmonic oscillator potential well,
the moment of inertia has in the limit of vanishing angular velocity exactly
the rigid-body value in any combination of stationary single-particle states
provided the total energy is stationary with respect to volume-conserving
variations of the equipotential ellipsoids~\cite{bm55}. At finite angular
velocity, the condition for the rigid-body value is that the second moments of
the density distribution should have the ratio of the squares of the axes of
the oscillator-plus-centrifugal equipotential ellipsoids, which is not exactly
equivalent to a stationary energy~\cite{Val}. These results have led to the
expectation that in real nuclei, the moment of inertia would not be very
different from the rigid-body value if the pairing is quenched. This
expectation was substantiated by early studies like, for example, that in
Ref.\cite{fp} of more realistic single-nucleon potentials, which seemed to
indicate that permitting the nuclear system to relax to an equilibrium shape
generally tends to reduce deviations from the rigid-body moment of inertia due
to shell structure. The validity of the aforementioned conjecture for the real
nuclear potential remains, however, a continuing subject of investigation with
new theoretical and experimental techniques, and so does the related question
of the current distribution in a rotating nucleus~\cite{Rad76,Kunz83,Kunz84}.

\begin{figure}

\psfig{file=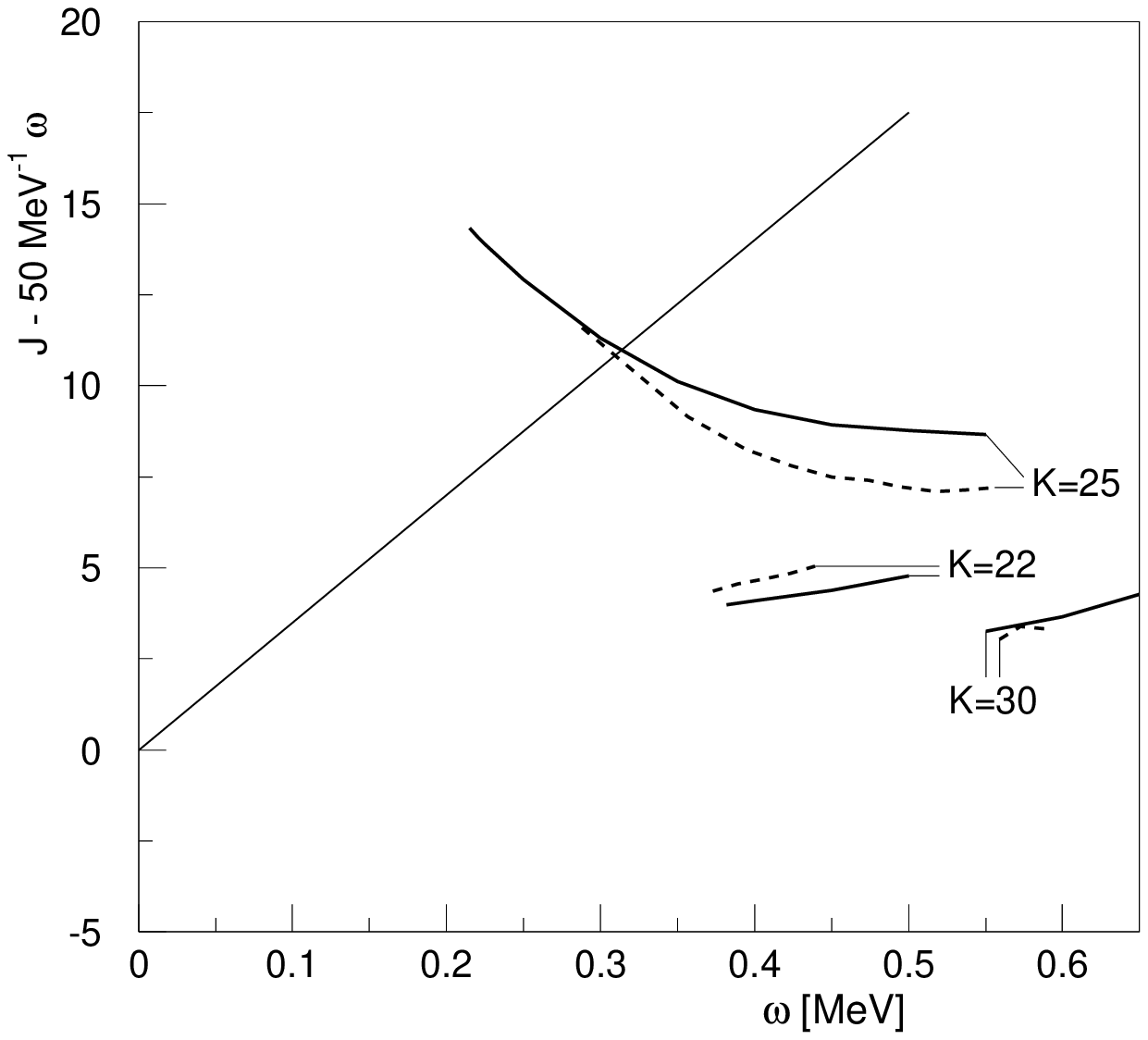,width=8cm}
\psfig{file=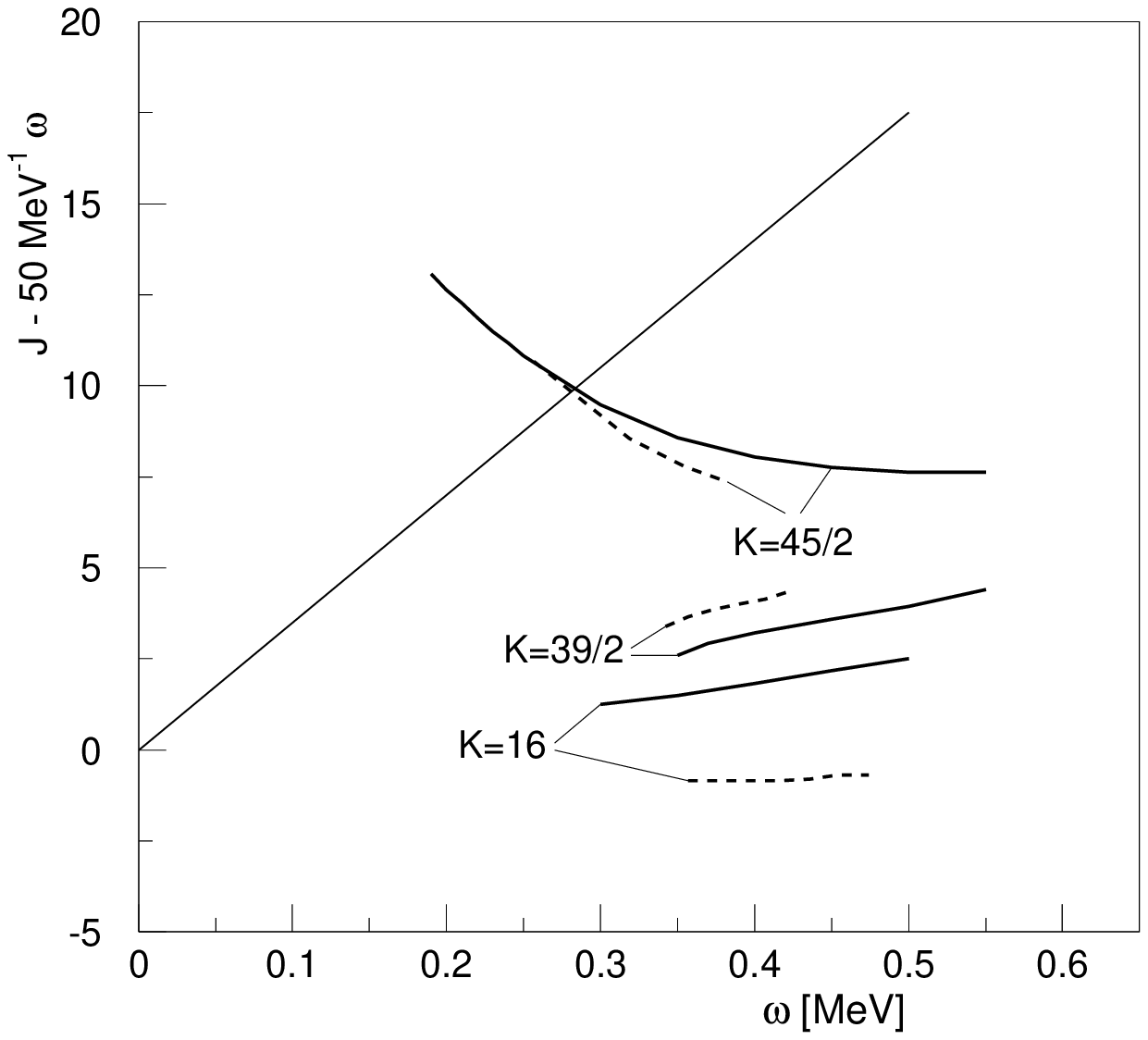,width=8cm}

\caption[]{\label{unpaired}
Functions $J(\omega)$ calculated without pairing for several
configurations listed in Table~\ref{table}. The value of $J$ is given relative
to a linear reference 50~MeV$^{-1}\omega$. Solid line: TAC calculation. Dashed
line: experiment~\cite{hf178,w178,Cu99,w179}. Upper panel: bands in~$^{178}$W.
Lower panel: bands in~$^{178}$Hf and~$^{179}$W. Moments of inertia can be
obtained from the graphs, as $J/\omega$ (kinematic value) and $dJ/d\omega$
(dynamic value). The straight line corresponds to the rigid moment of inertia
${\cal J}=85~\text{MeV}^{-1}$.}

\end{figure}

Systematic deviations from rigid-like behavior at high angular velocity have
been demonstrated for transitional nuclei in the A$\approx$110
region~\cite{Wa96} and discussed for superdeformed nuclei~\cite{Ra87,Be88}. In
the present study we address the inertial behavior of a different class of
nuclear excitations: high-seniority states in well deformed A$\approx$180
nuclei (see also the earlier work by Andersson~et~al.~\cite{An81}). Recently,
rotational bands have been observed in, for example, $^{178}$Hf, $^{178}$W
and~$^{179}$W~\cite{hf178,w178,Cu99,w179} that are built on configurations with
up to four broken pairs (that is, seniority eight) and high $K$ values, where
$K$ is the angular momentum with respect to the body-fixed deformation axis. It
is found (see the empirical data in Fig.~\ref{unpaired}) that the moments of
inertia are substantially below the rigid-body value. Furthermore, some bands
deviate from the linear dependence of the angular momentum on the angular
velocity, expected for the strong coupling of quasiparticles to the deformed
field.

These features can be explained~\cite{Dracoulis} by the persistence of pair
correlations in the Lipkin-Nogami pairing model, combined with the assumption
that the zero-pairing limit would result in rigid-like rotation. However, since
the latter assumption is not self-evident, a microscopic determination of the
moment of inertia in the zero-pairing limit is required. In the present work it
is demonstrated, through tilted-axis-cranking (TAC) calculations, that the main
experimental features can be understood by assuming that nucleons move in a
rotating mean field with {\em no pairing}. The preliminary results of
Ref.~\cite{fr95} are extended.

\section{The tilted-axis-cranking Model}\label{tac}

To describe the high-$K$ rotational bands, involving many unpaired nucleons and
predominant magnetic dipole transitions, the tilted-axis-cranking
approach~\cite{tac} is employed. When pairing is neglected, the nuclear state
$|\omega\rangle$ considered is a uniformly rotating Slater determinant which is
an eigenstate of the ``Routhian''
\begin{equation}\label{sp}
  h'=h_{\text{def}}(\varepsilon_2,\varepsilon_4)
  -\omega(j_1\sin\vartheta+j_3\cos\vartheta)\,,
\end{equation}
where $h_{\text{def}}$ is the Hamiltonian of independent nucleons in a deformed
potential, $\omega$ is the angular velocity, $j_1$ and $j_3$ the components of
the angular momentum with respect to the principal axes 1 and 3 (symmetry axis)
of the deformed potential, and $\vartheta$ the angle of the angular velocity
with the 3-axis. The total Routhian $E'$ is obtained by applying the Strutinsky
renormalization to the energy of the non-rotating system $E_0$. This kind of
approach has turned out to be a quite reliable calculation scheme in the case
of standard cranking~\cite{tbengtsson}. Thus we have
\begin{equation}\label{E'str}
  E'(\omega,\vartheta,\varepsilon_2,\varepsilon_4)
  =E_{LD}(\varepsilon_2,\varepsilon_4)-\tilde E
  +\langle\omega|h'|\omega\rangle\,.
\end{equation}
By means of Strutinsky-averaging~\cite{mo}, the smooth energy $\tilde E$ is
calculated from the non-rotating single-nucleon energies, obtained from the
Hamiltonian $h_{\text{def}}(\varepsilon_2,\varepsilon_4)$.

The orientation angle $\vartheta$ is found by requiring the total angular
momentum $\vec J=\langle\omega|\vec j|\omega\rangle$ to be parallel to
$\vec\omega$. This makes $E'$ a minimum with respect to $\vartheta$. In the
case of the high-K bands we are interested in, the rotational axis is
``tilted'', i. e. it does not coincide with one of the principal axes of the
deformed potential ($\vartheta\ne90^{\circ}\text{ or }0^{\circ}$). The
equilibrium shape is found by minimizing $E'$ with respect to the deformation
parameters $\varepsilon_2$ and $\varepsilon_4$ of the potential.

The calculated angular momentum $J(\omega)=\sqrt{J_1^{\,2}+J_3^{\,2}}$ is
compared with the corresponding 
experimental function, which is constructed by the standard
procedure: In terms of
the energy levels $E(I)$ of a $\Delta I=1$ rotational band, where $I$
denotes the angular momentum quantum number, one sets
$\omega(J)=E(I)-E(I-1)$ for $J=I$. For
a given observed band, this defines a discrete set of empirical pairs of $J$
and $\omega$ from which the experimental function $J(\omega)$ is obtained by
interpolation. (Taking $\omega(J)$ at $J=(I-\frac{1}{2})+\frac{1}{2}=I$
simulates an RPA-correction to the Hartree-Fock energy~\cite{Mar}).

\section{Single-nucleon Hamiltonian and deformations}\label{details}

In the present calculation for the nuclei $^{178}$Hf, $^{178}$W and~$^{179}$W,
the modified oscillator form~\cite{mo} of the Hamiltonian $h_{\text{def}}$ was
adopted. For the combinations of single-nucleon states listed in
Table~\ref{table}, the equilibrium shape at zero angular velocity was
\begin{table*}

\caption[]{Configurations and pair gaps for $\omega=0$ of the rotational bands
discussed in this paper. The states are labeled by their angular momentum $K$
with respect to the 3-axis and their parity $\pi$. Indicated is the composition
relative to the $^{178}$Hf or~$^{178}$W ground state in the absence of pairing,
as well as the contribution to $K^\pi$ of each kind of nucleon. The
$K^\pi=16^+$ band belongs to the nucleus~$^{178}$Hf, the $K^\pi=7^+$, $15^+$,
$22^-$, $25^+$ and~$30^+$ bands to the nucleus~$^{178}$W, and the
$K^\pi=39/2^+$ and~$45/2^-$ bands to the nucleus~$^{179}$W. The orbitals are
identified by their contribution to $K^\pi$. Holes are understood to occupy the
time-reversed orbital. The asymptotic quantum numbers are for the proton
orbitals: [514~9/2], [404~7/2], [541~1/2], [402~5/2], [505~11/2], and for the
neutron orbitals: [514~7/2], [633~7/2], [642~9/2], [512~5/2]. Note that the
$K^\pi=1/2^-$ proton orbital intrudes from the $1h_{9/2}$ spherical level. For
each kind of nucleon, the BCS and, in a bracket, Lipkin-Nogami pair gaps
calculated for $\omega=0$ are given.}
\begin{tabular}{|ccccc|}
K$^\pi$&proton configuration&$\Delta_p$ (MeV)&
neutron configuration&$\Delta_n$ (MeV)\\
\hline
$7^-$&$[\,]_{0^+}$&1.13&$[7/2^-,(7/2^+)^{-1}]_{7^-}$&0.48\\
$15^+$&$[9/2^-,(7/2^+)^{-1}]_{8^-}$&0(0.75)&
$[7/2^-,(7/2^+)^{-1}]_{7^-}$&0.48\\
$39/2^+$&
$[9/2^-,(7/2^+)^{-1}]_{8^-}$&0(0.75)&
$[9/2^+,7/2^-,(7/2^+)^{-1}]_{23/2^-}$&0(0.75)\\
$22^-$&$[9/2^-,(7/2^+)^{-1}]_{8^-}$&0(0.75)&
$[9/2^+,7/2^-,(5/2^-,7/2^+)^{-1}]_{14^+}$&0(0.51)\\
$45/2^-$&
$[1/2^-,9/2^-,(5/2^+,7/2^+)^{-1}]_{11^+}$&0(0.66)&
$[9/2^+,7/2^-,(7/2^+)^{-1}]_{23/2^-}$&0(0.75)\\
$25^+$&
$[1/2^-,9/2^-,(5/2^+,7/2^+)^{-1}]_{11^+}$&0(0.66)&
$[9/2^+,7/2^-,(5/2^-,7/2^+)^{-1}]_{14^+}$&0(0.51)\\
$30^+$&
$[11/2^-,9/2^-,(5/2^+,7/2^+)^{-1}]_{16^+}$&0&
$[9/2^+,7/2^-,(5/2^-,7/2^+)^{-1}]_{14^+}$&0\\
$16^+$&$[9/2^-,(7/2^+)^{-1}]_{8^-}$&0(0.84)&
$[7/2^-,(9/2^+)^{-1}]_{8^-}$&0(0.77)
\end{tabular}

\label{table}

\end{table*}
determined. Most of the configurations in $^{178}$W and~$^{179}$W were found to
have equilibrium values of the quadrupole deformation $\varepsilon_2$ and
hexadecapole deformation $\varepsilon_4$ (see~Ref.~\cite{mo}) close to
$\varepsilon_2=0.23$ and $\varepsilon_4=0.02$. Only the $K^{\pi}=45/2^-$
and~$25^+$ configurations, which have a proton in the $1h_{9/2}$ state, have
somewhat larger equilibrium deformations, given approximately by
$\varepsilon_2=0.25$ and~$\varepsilon_4=0.015$. In the $K^{\pi}=16^+$
configuration in $^{178}$Hf, the equilibrium shape has $\varepsilon_2=0.22$
and~$\varepsilon_4=0.05$. These values of the shape parameters were used
in the following calculations. The difference between the deformation of the
$K^{\pi}=45/2^-$ and~$25^+$ configurations and that of the other configurations
in $^{178}$W and~$^{179}$W changes the rigid-body moment of inertia by 6\,\%.

For the $K^{\pi}=45/2^-$ and~$25^+$ configurations, we studied the change of
equilibrium shape as a function of the angular velocity. In the relevant
interval of $\omega$, the variation of $\varepsilon_2$ stays below 0.005 and
that of $\varepsilon_4$ is negligible. This corresponds to a 2\,\% variation of
the rigid-body moment of inertia.

\section{Results and discussion}\label{results}

Figure~\ref{unpaired} shows the calculated and empirical functions $J(\omega)$
for the configurations in Table~\ref{table} except those with $K^{\pi}=7^-$
and~$15^+$. A close correspondence between calculation and data is apparent
from this figure. This includes recent data for a $K^{\pi}=30^+$ band in
$^{178}$W~\cite{Cu99}. It is also evident that the moments of inertia are
considerably smaller than the rigid-body value, which is about 85~MeV$^{-1}$
for these masses and shapes. The typical empirical moment of inertia is about
55~MeV$^{-1}$. The $K^{\pi}=45/2^-$ and~$25^+$ bands are discussed later.

 \begin{figure}

\psfig{file=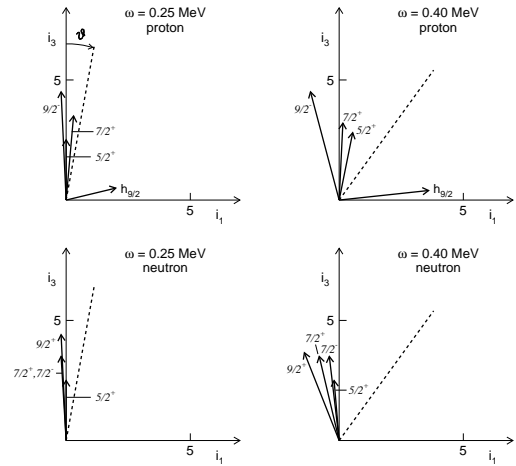,width=8cm}

\caption[]{The vectors $(i_1,i_3)=\langle(j_1,j_3)\rangle$ of the active
particles in the $K^\pi=25^+$ band. They are calculated for the eigenstates of
the Routhian~(\ref{sp}) without pairing at $\omega=0.25$~MeV and
$\omega=0.40$~MeV and the corresponding tilt angle $\vartheta$. For each vector
the label $K^\pi$ corresponds to the one in Table~\ref{table}, except that
$h_{9/2}$ corresponds to $1/2^-$ in the table. The dashed line shows the common
direction of the vectors $\vec\omega$ and~$\vec J$, which is tilted by the
angle $\vartheta$ from the 3-axis.}

\label{vectors}

\end{figure}

This strong deviation from the behavior of the moment of inertia in the limit
of large particle number~\cite{bm2,ring} may be understood from the details of
the shell structure at prolate deformation. Thus, the upper and middle parts of
the 50--82 proton and 82--126 neutron shells, where the Fermi levels are
situated in these nuclei (with $Z=72,\,74$ and $N=104\text{--}106$), have a
concentration of orbitals that are strongly coupled to the deformed potential.
This inhibits the generation of total angular momentum by alignment of the
angular momenta of the individual nucleons with the 1-axis. The result is a
moment of inertia that is smaller than the average. The effect is illustrated
by Fig.~\ref{vectors}, which shows that in comparison with the weakly coupled
$1h_{9/2}$ proton orbital, the angular momentum of the strongly coupled
orbitals tends to stay closely aligned with the 3-axis, and for some orbitals
even slightly antialigned with the 1-axis.

In contrast, moments of inertia above the average are expected for nuclei with
Fermi levels in the lower parts of the major shells. Such a variation was
actually found through the 82--126 neutron shell in the detailed calculations
in Ref.~\cite{fp}. With increasing deformation, the shell structure, and hence
its contribution to the moment of inertia, is progressively damped~\cite{fp}.
Less pronounced deviations from the rigid-body value are therefore expected for
superdeformed nuclei.

It should be noted that the substantial deviations from the rigid-body moment
of inertia seen in Fig.~\ref{unpaired} occur at the calculated equilibrium
shape of each configuration. A similar experience applies to the magnetic
susceptibility of small metal clusters~\cite{fpm}, which have a flat-bottom
single-particle potential like that of atomic nuclei. The deviation from the
rigid-body moment of inertia reflects a non-rigid flow of mass in the
rotational states. Such intrinsic mass currents have been discussed for atomic
nuclei by several authors~\cite{Rad76,Kunz83,Kunz84} as well as for small metal
clusters~\cite{fpm}.

The behavior of the $K^{\pi}=16^+$, $39/2^+$, $22^-$ and~$30^+$ bands is well
described in terms of a constant moment of inertia of each configuration with a
value about 55~MeV$^{-1}$. Such a constant moment of inertia corresponds to the
familiar expression for the energy levels in a rotational band built on a
strongly coupled intrinsic state, $E(I)=(I(I+1)-K^2)/2{\cal J}$, and it
indicates a collective origin of the angular momentum with respect to the
1-axis.

The $K^{\pi}=45/2^-$ and~$25^+$ bands show a totally different behavior with a
large up-curvature of the function $J(\omega)$. Asymptotically, in the limit of
large angular velocity, the moments of inertia approach values similar to those
of the other bands. As discussed in Ref.~\cite{Dracoulis}, this behavior
results from the presence of a $1h_{9/2}$ proton orbital in the configurations
of the $K^{\pi}=45/2^-$ and $25^+$ bands. In fact, as the component $\omega_1$
of the angular velocity becomes finite, this weakly coupled orbital, which
intrudes from the the $Z=82\text{--}126$ spherical shell, immediately aligns
its angular momentum with the 1-axis, thus making a significant contribution to
the component $J_1$ of the total angular momentum on the 1-axis. The situation
is illustrated in Fig.~\ref{vectors}.

\begin{figure}

\psfig{file=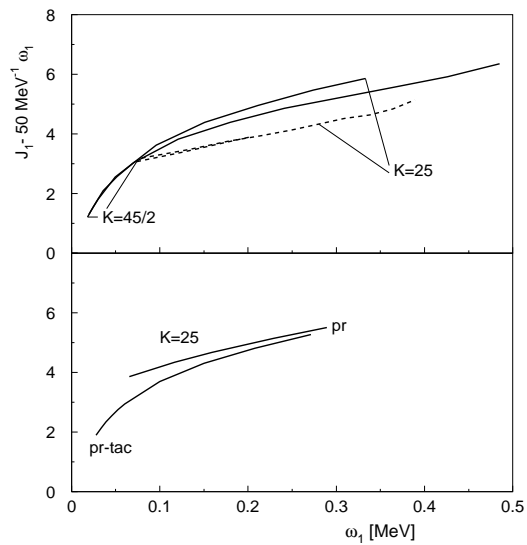,width=8cm}

\caption[]{Functions $J_1(\omega_1)$ for the $K^{\pi}=45/2^-$ and~$25^+$ bands.
The value of $J_1$ is given relative to a linear reference 50~MeV$^{-
1}\omega_1$.
The upper panel shows the experiment~\cite{w178,Cu99,w179} (dashed) and the
TAC-calculations (solid), which are without pairing. Note that these
calculations and data are the same as those in Fig.~\ref{unpaired}. They are
just presented differently. The lower panel shows the results of the schematic
model discussed in Sec.~\ref{pr}. Labels: pr-tac: cranking case, pr: quantal
case.}

\label{h9half}

\end{figure}

The functions $J_1(\omega_1)$ actually calculated for these two bands are shown
in Fig.~\ref{h9half}. Corresponding empirical functions were extracted from the
data by assuming, in close accordance with what is calculated, that $J_3$ is
constant and equal to $K$, i. e.
\begin{equation}\label{om1}
  J_1=\sqrt{J^2-K^2}\,,\quad\omega_1=\omega\sqrt{1-(K/J)^2}\,.
\end{equation}
In the empirical range of $\omega_1$, both the calculated and the measured
functions are seen to be fairly linear, and extrapolating these parts of the
curves to $\omega_1=0$ yields the common value $J_1=2.8\pm 0.5$
(cf.~Fig.~\ref{h9half}).

In order to see how the behavior of the $K^{\pi}=45/2^-$ and~$25^+$ bands seen
in Fig.~\ref{unpaired} may emerge from this picture, consider an idealized
scenario where the $1h_{9/2}$ proton orbital makes a constant contribution $i$
to $J_1$, and all orbitals together a constant contribution to $J_3$ equal to
$K$ and a contribution to $J_1$ equal to ${\cal J}_R{\omega_1}$, where
${\cal J}_R$ is a constant. (Such a schematic model is discussed in more detail
in Sec.~\ref{pr}). With $J_1={\cal J}_R\omega_1+i$ and (\ref{om1}), we have
\begin{equation}
  \omega=\left(1-\frac{i}{\sqrt{J^2-K^2}}\right)\omega_{\text{sc}}\,,\quad
  \omega_{\text{sc}}=\frac{J}{{\cal J}_R}\,,
\end{equation}
whence $\omega$ is seen to become smaller than the frequency
$\omega_{\text{sc}}$ for strong coupling ($i=0$).

In the calculation, there is a gradual increase of the contribution $i$ to
$J_1$ of the $1h_{9/2}$ proton orbital towards it maximum 9/2. Thus, the
assumption above of a constant $i$ was too schematic. The calculated curves
show a slight down-curvature due to saturation of $i$. The absence of a similar
down-curvature in the data might be the result of a counteractive non-linearity
of the remaining, collective, part of $J_1$. In that case, the present
calculation does not get this part quite right and overestimates the collective
moment of inertia by about 5~MeV$^{-1}$. Nevertheless, these considerations
show that (i) the essential difference of behavior, induced by the alignment
with the 1-axis of the angular momentum of the $1h_{9/2}$ proton, can be well
understood, and (ii) the collective part ${\cal J}_R$ is about 55~MeV$^{-1}$,
like for the other bands.

\section{Additional investigations}\label{additional}

The calculations for zero pairing, and their comparison with experimental data,
constitute the principal outcome of this work. However, it is also instructive
to investigate some finite-pairing effects and other model assumptions.

\subsection{Static pairing}\label{static}

Pairing is taken into account~\cite{tac} by including the pair field in the
quasiparticle Routhian
\begin{equation}\label{qp}
  h'=h_{\text{def}}+\Delta(P^++P)-\lambda N
  -\omega(j_1\sin\vartheta+j_3\cos\vartheta)\,,
\end{equation}
where $P^+$ is the monopole pair operator and $N$ is the particle number. In
order to keep the notation simple we do not distinguish between the proton and
neutron parts of the pair field. The rotating deformed state is obtained by
replacing the Slater determinant by the quasiparticle configuration
$|\omega\rangle$, which is the eigenstate of (\ref{qp}). The vector $\vec J\,$
is equal to $\langle\omega|\vec j|\omega\rangle$ with this new state
$|\omega\rangle$. The chemical potential $\lambda$ is fixed by requiring
$\langle\omega|N|\omega\rangle$ to be equal to the actual particle number, and
the pair gap $\Delta$ by the self-consistency condition
$\Delta=G\langle\omega|P|\omega\rangle$. For $\Delta=0$, this formalism is
equivalent to the previous one.

The pairing force constants $G_n$ and $G_p$ were determined by the condition
that the pair gaps in the nuclear ground state should be equal to the empirical
odd-even mass differences. It is well known from previous studies (for
instance, Ref.~\cite{shimizu}) that with increasing angular velocity, the pair
gaps and chemical potentials change their values essentially stepwise with a
successive breaking of Cooper pairs. Since a detailed description of the paired
state is not the concern of this paper, the chemical potentials and pair gaps
were kept constant for each configuration as long as there was no pair breaking
encountered.

The pair gaps determined at the band heads are listed in Table~\ref{table}. For
most of the configurations, they are seen to vanish. Exceptions are the
$K^{\pi}=7^-$ and~$15^+$ states. These have a common neutron configuration with
one broken Cooper pair, which leaves a reduced but finite neutron pair gap. The
$K^{\pi}=7^-$ state furthermore has the ground-state proton configuration and
hence the ground-state proton pair gap. The proton configuration of the
$K^{\pi}=15^+$ state is found in the calculation to be just on the border of
having a static proton pair field. Small variations of $G_p$ about the value
obtained by adjustment to the odd-even mass difference in fact cause $\Delta_p$
to vary between 0 and~0.5~MeV. For the calculations, we have chosen
$\Delta_p=0$, as also listed in Table~\ref{table}. This gives a good agreement
with the measured function $J(\omega)$.

\begin{figure}

\psfig{file=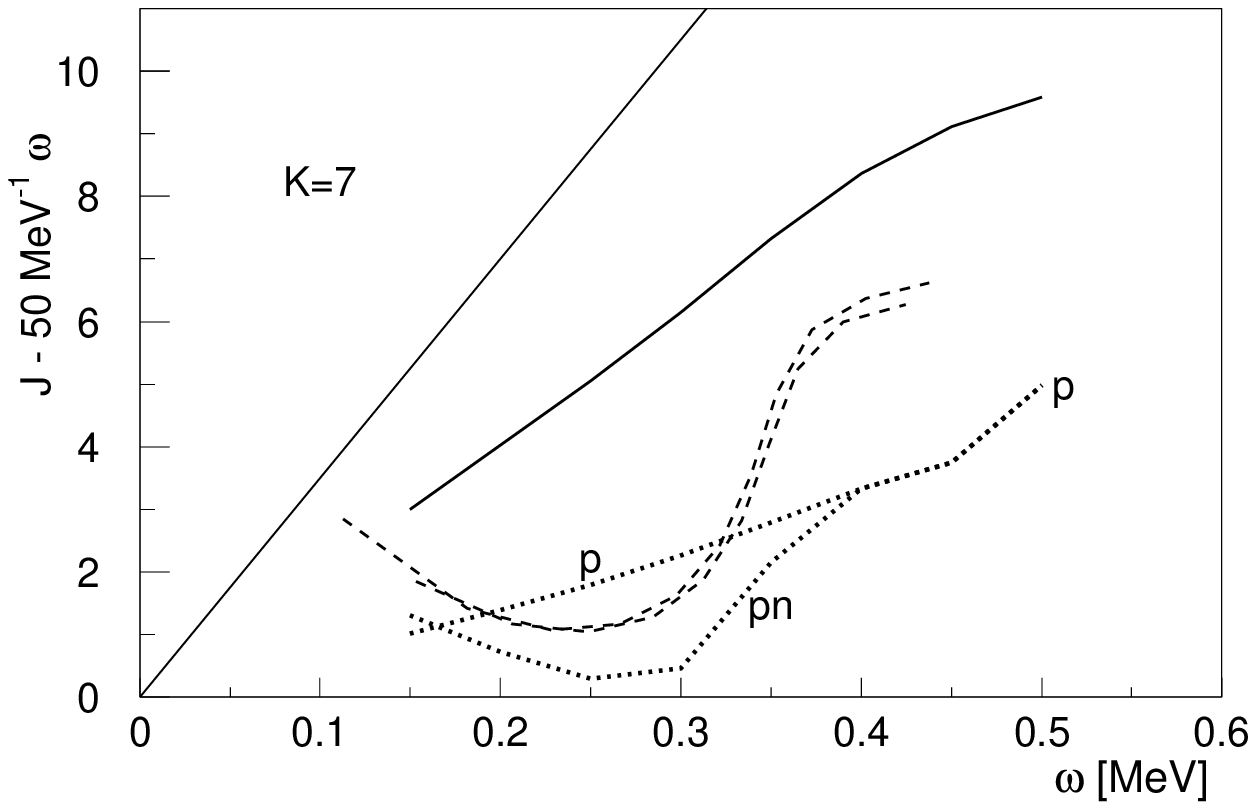,width=8cm}
\psfig{file=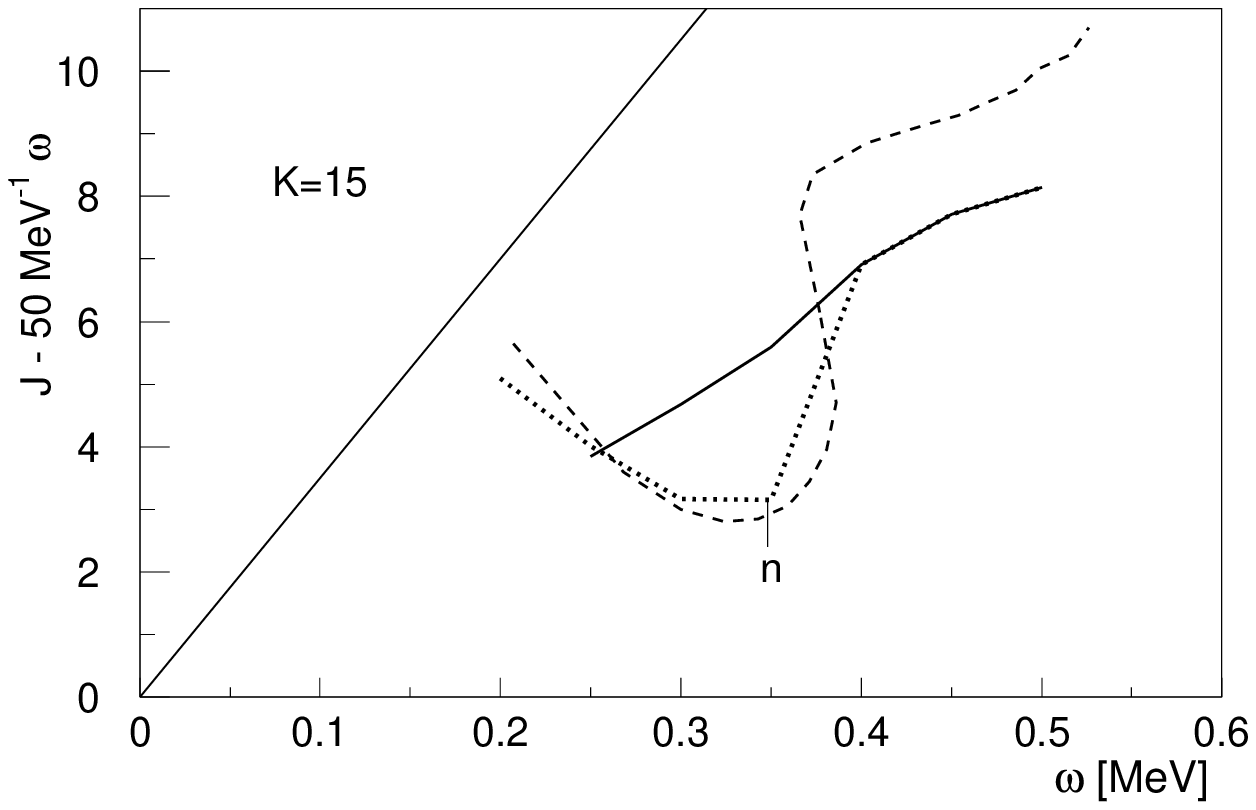,width=8cm}

\caption[]{Functions $J(\omega)$ calculated for the bands $K^\pi=7^-$ (upper
panel) and $K^\pi=15^+$ (lower panel) with and without static pairing. Solid
line: no pairing. Dotted line: with pairing. Dashed line:
experiment~\cite{w178,Cu99}. The labels additionally distinguish between
different combinations of the pair gaps: p:~$\Delta_p=1.13$~MeV, $\Delta_n=0$;
n:~$\Delta_p=0$, $\Delta_n=0.48$~MeV; pn:~$\Delta_p=1.13$~MeV,
$\Delta_n=0.48$~MeV. The curve n in the lower panel merges with the solid line
and the curve pn in the upper panel with the curve p because $\Delta_n=0$ is
found for $\omega\ge0.4$~MeV. The value of $J$ is given relative to a linear
reference 50~MeV$^{-1}\omega$. The straight line corresponds to the rigid
moment of inertia ${\cal J}=85~\text{MeV}^{-1}$.}

\label{bcs}

\end{figure}

Figure~\ref{bcs} shows the functions $J(\omega)$ calculated for the
$K^{\pi}=7^-$ and~$15^+$ bands. Both of them are seen to bend upwards near
$\omega=0.35$~MeV. This is because, by breaking a Cooper pair, two neutrons in
$1i_{13/2}$ orbitals align their angular momenta with the 1-axis. For
$\omega\ge0.4$~MeV, a vanishing pair gap is calculated for this neutron
configuration. Therefore, we let in the figure the curves calculated with the
neutron pair gap at the band head join for $\omega\ge0.4$~MeV those calculated
with $\Delta_n=0$. These are about 2 units below the measured curves in this
range of $\omega$. We could not find a reason for the discrepancy.

This pair breaking is of the type known as a $BC$-crossing (see, for example,
Ref.~\cite{csm}). As also seen from Fig.~\ref{bcs}, no similar upbends arise in
the case $\Delta_n=0$. This conforms to the general experience~\cite{stapair}
that a static pair field is required for band crossings of the types $AB$,
$BC$, etc.. Thus, the presence of upbends in the data is evidence for a static
neutron pair field in these bands.

\subsection{Pair fluctuations}\label{fluct}

Near the critical point of the vanishing of the static pair gap, large
fluctuations of the pair field, known as dynamic pair correlations, are
expected~\cite{shimizu,diebel}. Dynamic pair correlations are taken into
account in an approximate way by the Lipkin-Nogami correction for the
fluctuation of particle number in the BCS state. (See Ref.~\cite{lipkin} and
refs. therein). For several configurations, we made the Lipkin-Nogami
calculation at the band head. The resulting Lipkin-Nogami pair gaps are also
shown in Table~\ref{table}. With these gaps, $J(\omega)$ was calculated as in
the case of static pairing (see Sec.~\ref{static}), except that the chemical
potentials were adjusted with the angular velocity so as to keep the correct
expectation values of the proton and neutron numbers.

\begin{figure}

\psfig{file=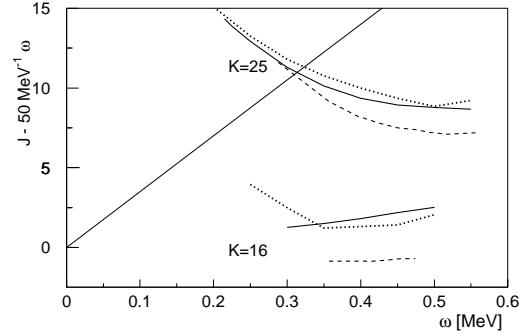,width=8cm}

\caption[]{Influence of the dynamical pair correlations on $J(\omega)$ for the
$K^{\pi}=16^+$ and~$25^+$ bands. Solid line: no pairing. Dotted line:
calculated using the Lipkin-Nogami pair gaps quoted in Table 1. Dashed line:
experiment~\cite{hf178,w178,Cu99}. The value of $J$ is given relative to a
linear reference 50~MeV$^{-1}\omega$. The straight line corresponds to the
rigid moment of inertia ${\cal J}=85~\text{MeV}^{-1}$.}

\label{ln}

\end{figure}

The calculated functions $J(\omega)$ for the $K^\pi=16^+$ and~$25^+$ bands
shown in Fig.~\ref{ln} are representative for the results. It is seen that
relative to the calculation without pairing, the pair fields produced by the
Lipkin-Nogami pair gaps make only minor corrections to the angular momentum (of
the order of 1 unit), which do not improve the agreement with experiment. Thus,
pair fluctuations appear to be inessential for the explanation of the observed
deviations from the rigid value of the moments of inertia at high values of
$K$.

This result  may seem to be
 at variance with the investigation of low-$K$ bands in
Refs.~\cite{shimizu,diebel}. There it was found that at frequencies where the
static pair gap is zero the pair fluctuations reduce the angular momentum by
3--4 units in the yrast band of even-even nuclei. The different sensitivity to
the pair correlations may be understood. In order to generate the angular
momentum along the 3-axis (high $K$) several pairs are broken. This blocks the
affected single particle states from taking part in the pair correlations.
However, it is just the contribution of these particles near the Fermi surface
which is most sensitive to the pair correlations. In the case of the yrast
bands of the even-even nuclei only one neutron pair (1$i_{13/2}$) is broken.
Consequently these bands are more sensitive to the pair fluctuations. This
argument is consistent with Refs.~\cite{shimizu,diebel}, where it was found
that in bands with two broken pairs (odd-$A$ nuclei and negative parity bands
in even-even nuclei) the pair fluctuations reduce the angular momentum only by
1--2 units. Hence, only the low-$K$ bands are suited to study the influence of
the pair fluctuations on the moments of inertia.

\subsection{A particle-rotor model calculation}\label{pr}

It was seen in Sec.~\ref{results} that the behavior of the $K^{\pi}=45/2^-$
and~$25^+$ bands at low angular velocity is largely determined by a single
proton in a $1h_{9/2}$ orbital. The behavior was qualitatively explained in
terms of a particle-rotor model where all nucleons except the $1h_{9/2}$ proton
are assumed to make a constant contribution to $J_3$ equal to
$K_R=K-\frac{1}{2}$ and a contribution to $J_1$ equal to ${\cal J}_R\omega_1$,
where ${\cal J}_R$ is a constant. This situation may be further analyzed by
calculating the quantal states of this model. In particular, we address the
question of whether the deviation between the experiment and the calculation in
the upper panel of Fig.~\ref{h9half} are related to the violation of angular
momentum conservation in the TAC model. A quantal treatment of the system of a
particle coupled to a $K_R\ne0$ rotor was given previously in
Ref.~\cite{neergard}.

The coupling of the $1h_{9/2}$ proton to the deformed core is treated in a
schematic way. The particle space is restricted to a multiplet of
angular-momentum eigenstates with quantum number $j=9/2$, and $h_{\text{def}}$,
acting on the single proton, is taken to be a quadratic function of $j_3$. The
coefficient of this quadratic function is chosen so as to reproduce the
splitting of the $1h_{9/2}$ proton level found for the full Hamiltonian
$h_{\text{def}}$ at the deformations of the $K^{\pi}=45/2^-$ and~$25^+$ band
heads (see Sec.~\ref{details}).

The particle-rotor problem can be treated in the semiclassical TAC
approximation. The details are described in Ref.~\cite{fm}. The function
$J_1(\omega_1)$ of the $K^{\pi}=25^+$ band thus calculated with
${\cal J}_R$=55~MeV$^{-1}$ is shown in the lower panel of Fig.~\ref{h9half}. It
is seen that the schematic model reproduces the result of the full TAC
calculation, seen in the upper panel, very closely.

The result of the exact quantal treatment of the same particle-rotor model is
also shown in the lower panel of Fig.~\ref{h9half}. In order to generate the
plot the quantal energies are treated like empirical ones (see Secs.~\ref{tac}
and~\ref{results}). The quantal calculation conforms better to the data than
the TAC approximation in producing a more linear function $J_1(\omega_1)$.
However, extrapolating this function from the empirical range of $\omega_1$ to
$\omega_1=0$ yields $J_1=3.5$, which is significantly larger than the empirical
value $J_1=2.8$.

The different behaviors of the quantal particle-rotor model and the TAC
approximation to it arise essentially from replacing the recoil energy
$(j_1^{\,2}+j_2^{\,2})/2{\cal J}_R$ by
$\langle j_1\rangle^2/2{\cal J}_R$~\cite{fm}. While the former is approximately
a constant, the latter acts as a potential that hinders the increase of
$\langle j_1\rangle$. Contrary to the quantal model, the cranking model was
seen to reproduce the extrapolated value of $J_1$ found empirically for the
$K^{\pi}=45/2^-$ and~$25^+$ bands. Thus, the nuclear system does not seem to
absorb the recoil angular momentum of the $1h_{9/2}$ proton into just a single
degree of freedom, as assumed in the quantal particle-rotor model. The present
study does not provide an answer to the interesting question: How can the
experimental curve $J_1(\omega_1)$ be so strikingly linear while the alignment
is far from being complete?

\begin{figure}

\psfig{file=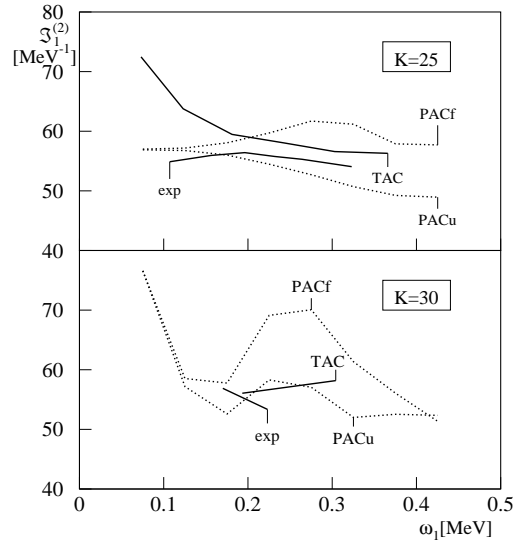,width=8cm}

\caption[]{Moments of inertia $dJ_1/d\omega_1$. Solid line: TAC calculation.
Dotted line: PAC calculation, favored band. Dotted and dashed line: PAC
calculation, unfavored band. Dashed line: experiment~\cite{w178,Cu99}. Upper
panel: $K^\pi=25^+$ band. Lower panel: $K^\pi=30^+$ band. The discrepancy
between the TAC calculation and the experimental data for the $K^\pi=25^+$ band
at low $\omega_1$ is discussed at the end of Sec.~\ref{results}.}

\label{pac}

\end{figure}

\subsection{How important is tilting the cranking axis?}\label{ral}

In the standard principal-axis cranking (PAC) model, $\omega_3=0$ is assumed.
Thus, one obtains a function $J_1(\omega_1)$. A corresponding empirical
function is extracted from the data by combining the TAC geometry with the
assumption $J_3=K=\text{constant}$ for a rotational band with a band head
angular-momentum quantum number $K$ \cite{bf}. What makes the essential
difference between the PAC and TAC models is thus the term $-\omega_3 j_3$ in
the Routhian~(\ref{sp}) of the latter. This term violates the invariance under
rotation by the angle $180^{\circ}$ about the 1-axis, whose eigenvalue is the
``signature''. In the PAC model, a ``favored'' and an ``unfavored'' function
$J_1(\omega_1)$, where the latter is the larger, are associated with a
configuration with $K\ne0$. These functions have opposite signature and
correspond to two separate level sequences with $\Delta I=2$.

Derivatives $dJ_1/d\omega_1$ for the $K^{\pi}=25^+$ and~$30^+$ bands calculated
in both models are compared with the corresponding empirical data in
Fig.~\ref{pac}. The derivative is seen to depend much more violently on
$\omega_1$ in the PAC model than in the TAC model. Furthermore, the PAC
calculation shows a substantial signature splitting. Since neither of these
features is seen in the data, it must be concluded that the term
$-\omega_3 j_3$ in the Routhian~(\ref{sp}) is significant for the description
of these high K-bands. The difference between the PAC and TAC results is larger
for the $K^{\pi}=30^+$ than the $K^{\pi}=25^+$ band. This is due to the smaller
deformation $\varepsilon_2$ of the former.

\section{Conclusion}\label{concl}

It has been shown quantitatively how the moments of inertia in the zero-pairing
limit may be substantially lower than the rigid-body value, indicating the
presence of mass currents of quantal origin in the body-fixed frame of
reference. Lower-than-rigid moments of inertia are both calculated and observed
systematically for rotational bands in $^{178}$Hf, $^{178}$W and~$^{179}$W,
where the neutron and proton Fermi levels are in the mid-to-upper portions of
their respective shells. The analysis of a number of high-seniority bands shows
that they behave as if the nuclei rotate in the unpaired state. The limited
sensitivity of the calculated multi-quasiparticle rotational motion to pair
gaps in the range 0--50\% of their full value suggests that moments of inertia
of high-$K$ bands may not be significantly affected by dynamic pair
correlations.

\acknowledgments

This work is supported in part by the UK Engineering and Physical Sciences
Research Council and by the DOE grant DE-FG02-95ER40934.

\end{document}